\documentclass[aps,pre,twocolumn,groupedaddress]{revtex4-1}
\usepackage{graphicx}
\usepackage{dcolumn}
\usepackage{bm}
\usepackage{amssymb}
\usepackage{color}

\begin{document}


\title{Angular-momentum conservation in discretization of Navier-Stokes equation for viscous fluids}

\author{Hiroshi Noguchi}
\email[]{noguchi@issp.u-tokyo.ac.jp}
\affiliation{
Institute for Solid State Physics, University of Tokyo,
 Kashiwa, Chiba 277-8581, Japan}

\date{\today}

\begin{abstract}
Although the Navier-Stokes equation (NSE) is derived under angular-momentum conservation (AMC),
numerical simulation methods often lack it. Here, we reveal that AMC violations result from 
implementation of the degenerated viscous terms of NSE.
To maintain AMC,
these degenerated terms must be separately integrated in accordance with their stress origins.
As observed in particle-based hydrodynamics methods, the violation causes artificial rotations 
in multi-component fluids with different viscosities.
At the interface between two fluids or with a mobile solid object,
AMC must be satisfied, whereas AMC can be neglected in bulk fluids.
We also clarify that the condition for constant fluid rotation 
as a rigid body in a container rotating at a constant speed is not the AMC of the stresses,
but the invariance of the viscous forces under a global rotation.
To confirm our theory,
we simulated the circular laminar flows of single- and binary-component fluids
using two-dimensional Lagrangian finite volume methods.
The results show excellent agreement with the analytical predictions
for fluids with and without AMC.
\end{abstract}


\maketitle

\section{Introduction}

Computer simulations have become one of the primary tools for investigating and 
visualizing fluid dynamics. Many types of simulation methods have been developed, but 
improvements are still continuously conducted. 
These methods are categorized into two groups. 
The first group directly discretizes the Navier-Stokes equation (NSE) or its variants,
which include the finite difference method (FDM), finite volume method (FVM), 
finite element method (FEM), and so on~\cite{ferz02,wend09}. 
The second group is based on a microscopic or mesoscopic equation of motion 
({\it e.g.}, Newton's equation of motion and the Boltzmann equation); 
these equations are different from NSE, although large-scale dynamics follow NSE.
In the second group, the lattice Boltzmann method \cite{succ01} and three particle-based methods
({\it i.e.}, direct simulation Monte Carlo (DSMC)\cite{bird98}, dissipative particle dynamics (DPD) 
\cite{groo97,espa17}, and multi-particle collision (MPC) dynamics \cite{kapr08,gomp09a})
 have been widely adopted.
These particle-based methods take thermal fluctuations into account and share the main constitution form,
while different implementations are employed for local viscous interactions~\cite{nogu07}.
These viscous interactions must conserve translational momentum to properly generate hydrodynamic flow, 
while angular-momentum conservation (AMC) can be switched on or off~\cite{nogu07}.
AMC is required for the calculation of torques and flows 
which boundary conditions are given by stress balances,
such as those found in multi-component fluids with different viscosities~\cite{goet07}.
AMC is also required to reproduce the friction of slip boundary conditions~\cite{yang15}.
However, the absence of AMC does not affect the velocity field of a fluid when the boundary conditions
are given by velocities. 
Thus, in the particle-based methods, AMC is explicitly imposed or
neglected depending on the application, and AMC's effects on these methods are well
understood.

Compared to the above particle-based methods,
AMC has not been well discussed in the discretized NSE methods.
Recently, the violation of AMC has been an issue in smoothed particle hydrodynamics (SPH)~\cite{mona05,pric12,elle18}.
SPH is a particle-based method, although the equation of motion is derived from
NSE in the Lagrangian frame of reference. 
Thus, SPH is positioned in the crossover area between the first and second groups. 
The original version of SPH did not include thermal fluctuations,
but  Espa{\~n}ol and Revenga developed smoothed dissipative particle dynamics (SDPD), a version of SPH including
thermal fluctuations~\cite{espa03}.
SDPD can also be considered as an extension of DPD. In conventional
SPH, the viscous forces do not conserve angular momentum.
To recover AMC in SDPD, Hu
and Adam~\cite{hu06} restricted the viscous forces only along the vectors 
between two interacting particles, as in DPD. 
In contrast, M{\"u}ller {\it et al.}~\cite{mull15} 
added an internal degree of freedom,
spin, from the fluid particle model~\cite{espa98} to the SDPD particle and applied NSE with the
spins to derive the particle motion. 
The rotational flow of binary
fluids and the vesicle dynamics in shear flow~\cite{mull15} and torque under the no-slip boundary condition~\cite{bian12} are properly reproduced by these AMC modifications of SDPD.
These results are well understood when the derived equations
of motion are viewed as one of the mesoscopic simulation methods described in Ref.~\cite{goet07}.
However, the results are unclear from the viewpoint of NSE discretization, because NSEs
are derived under the AMC condition. For example, Ellero and Espa{\~n}ol mentioned ``As the
SDPD equations are a discretization of NSEs which conserve angular momentum, angular
momentum is conserved in SDPD in the limit of high resolution'' in Ref.~\cite{elle18}. 
Thus, it has been considered that the lack of AMC is caused by a numerical error in the discretization,
although its mechanism is not known.

In this paper, we describe how AMC is lost in NSE discretization and how it can be
recovered. We found that the degeneracy in the derivatives of shear and volume stress terms
is the origin; as a result, the lack of AMC does not disappear in the limit of high resolution.
In Sec.~\ref{sec:theory}, we describe NSEs for fluids with and without AMC. Hereafter, we refer to fluids
with and without AMC as $+a$ and $-a$ fluids, respectively~\cite{nogu07}.
We show analytical solutions for the planar and rotational laminar flows of the $+a$ and $-a$ fluids. 
Next, in Sec.~\ref{sec:AMC}, we explain the AMC loss process 
using the Lagrangian finite volume method (LFVM)~\cite{serr01,spri10,duff11}.
LFVM is the most suitable method for this purpose: because it employs fewer approximations than
SPH, the mechanism is clearly captured. 
This mechanism is general and is shared by SPH and other methods. 
The details of LFVM in two-dimensional (2D) space are then described in Sec.~\ref{sec:LFVM2D}.
The method in Ref.~\cite{serr01,serr06} is 
extended to simulate the $+a$ fluids. The condition of the rigid-body rotation of fluids is
described in Sec.~\ref{sec:rbr}.
We found that it is determined by whether the discretized viscous forces are invariant 
for a global rotation, ${\bf v}={\boldsymbol \Omega}\times {\bf r}$.
The implementation of the boundary conditions and numerical details are described 
in Secs.~\ref{sec:bd} and \ref{sec:detail}, respectively.
In Sec.~\ref{sec:results}, 
the simulation results of the planar and rotational laminar flows of the $+a$ and $-a$
fluids are shown. The theoretical predictions given in Sec.~\ref{sec:ana}  are accurately reproduced
by these simulations. Sec.~\ref{sec:sum} contains our summary and discussion.

\section{Theory for fluids with and without angular-momentum conservation}\label{sec:theory}
\subsection{Navier-Stokes Equations}\label{sec:NSE}

In conventional viscous fluids that do conserve angular momentum, 
the stress is expressed by a symmetric tensor:
\begin{equation} \label{eq:strs+a}
\sigma_{\alpha\beta}^{(+)}= -P\delta_{\alpha\beta}
+ \eta \left(\frac{\partial v_{\alpha}}{\partial x_{\beta}}
+\frac{\partial v_{\beta}}{\partial x_{\alpha}} \right)  + \lambda(\nabla \cdot {\bf v})\delta_{\alpha\beta},
\end{equation}
where $\alpha,\beta \in \{x,y,z\}$ and P is the pressure.
The superscripts $(+)$ and $(-)$ are used to distinguish the quantities of the $+a$ and $-a$ fluids.
Here, $\eta$ is the viscosity and $\lambda$ is the second viscosity coefficient.
For the $-a$ fluids, the latter half of the shear stress, $\eta\partial v_{\beta}/\partial x_{\alpha}$, 
is removed as
\begin{equation} \label{eq:strs-a}
\sigma_{\alpha\beta}^{(-)}= -P\delta_{\alpha\beta}
+ \eta \frac{\partial v_{\alpha}}{\partial x_{\beta}} + \lambda(\nabla \cdot {\bf v})\delta_{\alpha\beta}.
\end{equation}
Note that the asymmetric stress tensor is generally expressed by
the addition of the asymmetric term
$\check{\eta} (\partial v_{\alpha}/\partial x_{\beta}- \partial v_{\beta}/\partial x_{\alpha} )$
to Eq.~(\ref{eq:strs+a}) as described in Ref.~\cite{goet07}.
However, in the simulation methods in which angular momentum is not conserved,
the asymmetric stress tensor typically has the form of Eq.~(\ref{eq:strs-a}).
This is because the asymmetric stress is introduced during simplification of the numerical procedure.
Introduction of asymmetric stress during the discretization of NSEs is described in the next section. 
In the $-a$ versions of MPC and DPD, the collisional stress follows Eq.~(\ref{eq:strs-a}), 
whereas the kinetic stress maintains the symmetry~\cite{nogu08}.

The velocity evolution equation ({\it i.e.}, NSE) of the $+a$ fluids is given by
\begin{eqnarray}
\rho\frac{D v_{\alpha}}{D t} &=& \frac{\partial \sigma_{\alpha\beta}^{(+)}}{\partial x_\beta} + f_{\rm ex,\alpha} \nonumber \\
&=& -\frac{\partial P}{\partial x_\alpha} 
+ \frac{\partial}{\partial x_\beta} \Big( \frac{\eta \partial v_{\alpha}}{\partial x_{\beta}}\Big)  \label{eq:nse0} \\ \nonumber
&& + \frac{\partial}{\partial x_\beta}  \Big(\frac{\eta \partial v_{\beta}}{\partial x_{\alpha}}\Big)  
+ \frac{\partial}{\partial x_\alpha} \Big(\frac{\lambda \partial v_{\beta}}{\partial x_{\beta}}\Big) + f_{\rm ex,\alpha},
\end{eqnarray}
where the Einstein notion is used,
 $\rho$ is the mass density, and $D/Dt$ is the Lagrangian derivative: $D/Dt=\partial/\partial t+ {\bf v}\cdot \nabla$.
The last term $f_{\rm ex,\alpha}$ expresses the $\alpha$ component of an external force ${\bf f}_{\rm ex}$, 
but hereafter we omit it in NSE for simplicity.
For the $-a$ fluids, the third term is removed.

When the viscosity coefficients $\eta$ and $\lambda$ are constant in the fluid,
the third and fourth terms of Eq.~(\ref{eq:nse0}) are degenerated as
\begin{equation} \label{eq:nse1}
\rho\frac{D {\bf v}}{D t} = -\nabla P + \eta \nabla^2 {\bf v} + \mu \nabla(\nabla\cdot{\bf v}),
\end{equation}
where $\mu=\eta+\lambda$ and $\mu=\lambda$ for the $+a$ and $-a$ fluids, respectively.
Interestingly, the derivatives of the asymmetric tensor 
$\partial v_{\beta}/\partial x_{\alpha}$  and the symmetric tensor ($\nabla\cdot{\bf v})\delta_{\alpha\beta}$
are identical.
Therefore, the $+a$ and $-a$ fluids follow the same equation;
when $\lambda^{(-)}= \eta+\lambda^{(+)}$ is used in the NSE of the $-a$ fluids,
no difference in fluid motion is detected.
In other words, Eq.~(\ref{eq:nse1}) by itself does not guarantee AMC.

The $-a$ fluids have an asymmetric stress that generates an artificial torque~\cite{goet07}.
In cylindrical coordinates ($r,\theta,z$), the azimuthal stress is given by
\begin{eqnarray} \label{eq:strs_rq+a}
\sigma_{r\theta}^{(+)} &=& \eta r
\frac{\partial \omega}{\partial r}, \\ \label{eq:strs_rq-a}
\sigma_{r\theta}^{(-)} &=& \eta r
\frac{\partial \omega}{\partial r}
+ \eta\omega,
\end{eqnarray}
for the $+a$ and $-a$ fluids, respectively,
where the angular velocity $\omega=v_{\theta}/r$.
The stress of the $+a$ fluids
depends only on the derivative of the angular velocity $\omega$.
In contrast, the $-a$ fluids have 
an additional term that is proportional to $\omega$.
Therefore, when a whole fluid slowly rotates as a rigid body,
a $-a$ fluid receives an artificial torque whereas a $+a$ fluid receives no torque. 
This additional torque modifies the torque balance on the interface between two fluids with different viscosities,
such that multi-component $-a$ fluids show different flow behavior than $+a$ fluids; this is described in the next subsection.

\subsection{Analytical Solutions of Laminar Flows}\label{sec:ana}

Since our interest is the viscous term,
we consider planar and circular laminar flows under a low Reynolds number and low Mach number.
The analytical solutions for planar and circular laminar flows are respectively described in Secs.~{\ref{sec:planar} and {\ref{sec:circular}.
Single-component and binary-component fluids are considered.
The viscosity of each component is constant.

\subsubsection{Planar Laminar Flows}\label{sec:planar}

A periodic force is imposed on fluids  in the $y$ direction 
as ${\bf f}_{\rm ex}(x)= f_0 \sin(kx) {\bf e}_y$ with the wavelength $L_x=2\pi/k$
where ${\bf e}_y$ is the unit vector along the $y$ axis.
The $+a$ and $-a$ fluids exhibit identical laminar flows.
The steady flow of the single-component fluid is given by ${\bf v}(x)=(f_0/\eta k^2)\sin(kx){\bf e}_y$~\cite{serr06}.
When the initial velocity is given by ${\bf v}(x)=v_{\rm a}(0)\sin(kx){\bf e}_y$,
the velocity field develops as
\begin{eqnarray}
 v_y(x,t) &=& v_{\rm a}(t)\sin(kx), \qquad v_x=v_z=0,\\
v_{\rm a}(t) &=& \frac{f_0}{\eta k^2}+ \Big( v_{\rm a}(0)- \frac{f_0}{\eta k^2} \Big)\exp\Big(-\frac{\eta k^2}{\rho}t\Big).
 \label{eq:sin1}
\end{eqnarray}

For binary fluids, the velocity $v_y$ in the steady flow is given by
\begin{eqnarray}
 \label{eq:sin2}
v_y(x) &=& v_{\rm a}\sin(kx) + v_{\rm s}, \\ \nonumber
v_{\rm a}&=& \left\{
\begin{array}{ll} f_0/\eta_0 k^2 & {\rm for} \quad n-1/2<x/L_x\le n, \\
                 f_0/\eta_1 k^2 & {\rm for} \qquad \qquad n <x/L_x\le n+1/2,
\end{array}
\right.
\end{eqnarray}
where $v_{\rm s}= (1/\eta_0-1/\eta_1)f_0/\pi k^2$ for a flow in which
 total translational momentum is zero.
Two fluids have the viscosities $\eta_0$ and $\eta_1$ and 
fill in the regions of $n-1/2<x/L_x\le n$ and $n <x/L_x\le n+1/2$, respectively, where $n$ is an arbitrary integer.

\subsubsection{Circular Laminar Flows}\label{sec:circular}

The fluids filling a single cylinder and the space between two concentric cylinders are considered 
as shown in Fig.~\ref{fig:cyl}(a) and (b), respectively.
The azimuthal velocity in a steady rotational flow is given by \cite{trit88,goet07}
\begin{equation} \label{eq:v_cf}
v_\theta(r) = A_ir+ \frac{B_i}{r},
\end{equation}
where $i=0$ and $i=1$ for the outer and inner fluids, respectively.
The $+a$ fluids in the single cylinder
rotate with a constant angular velocity (similar to that of a rigid body), {\it i.e.}, 
\begin{equation}
A_0=A_1=\Omega_{\rm out}, \qquad B_0=B_1=0,
\end{equation}
which is independent of the viscosity ratio $\eta_1/\eta_0$,
because the rigid-body rotation yields no gradient of the angular velocity $\partial \omega/\partial r$.
However, the $-a$ fluids have a stress term proportional to the absolute angular velocity $\omega$ [see Eq.~(\ref{eq:strs_rq-a})];
thus, the stress balance at the fluid boundary is expressed as
$(\eta_1-\eta_0)\omega(R_{\rm m})= -2\eta_0B_0/{R_{\rm m}}^2$.
Therefore, the flow is given by Eq.~(\ref{eq:v_cf}) with
\begin{eqnarray}
A_0 &=& \frac{\Omega_{\rm out}}{1-\frac{\eta_1-\eta_0}{\eta_1+\eta_0}\frac{{R_{\rm m}}^2}{{R_{\rm out}}^2}} , \quad B_0= - \frac{{R_{\rm m}}^2\Omega_{\rm out}}{\frac{\eta_1+\eta_0}{\eta_1-\eta_0}-\frac{{R_{\rm m}}^2}{{R_{\rm out}}^2}} ,\nonumber \\
A_1 &=&  \frac{\frac{2\eta_0}{\eta_1+\eta_0}\Omega_{\rm out}}{1-\frac{\eta_1-\eta_0}{\eta_1+\eta_0}\frac{{R_{\rm m}}^2}{{R_{\rm out}}^2}} ,    \quad B_1=0.
\label{eq:sng-a}
\end{eqnarray}

In the  circular Couette flow between two concentric cylinders, 
the coefficients $A_i$ and $B_i$ are obtained from the boundary conditions at $R_{\rm m}$,  $R_{\rm in}$, and $R_{\rm out}$.
For the $+a$ fluids, they are given by
\begin{equation}
B_0 = \frac{\Omega_{\rm in} - \Omega_{\rm out}}{ \frac{\eta_0}{\eta_1}\frac{1}{{R_{\rm in}}^2} - \frac{1}{{R_{\rm out}}^2} + \big(1-\frac{\eta_0}{\eta_1}\big)\frac{1}{{R_{\rm m}}^2} }, \quad B_1= \frac{\eta_0}{\eta_1}B_0, 
 \label{eq:dw+a}
\end{equation}
and 
\begin{equation}
A_0 = \Omega_{\rm out} -  \frac{B_0}{{R_{\rm out}}^2}, \qquad  A_1 = \Omega_{\rm in} -  \frac{B_1}{{R_{\rm in}}^2}.
 \label{eq:dw_cfa}
\end{equation}
Thus, the constant torque $4\pi\eta_0B_0$ propagates from the inner to outer cylinders via the fluids
because of AMC.
However, in the $-a$ fluids, the lack of AMC generates an additional torque, and the flow is modified as
\begin{eqnarray} \label{eq:dw-a}
B_0 &=& \frac{ \frac{2\eta_1}{\eta_1+\eta_0} {R_{\rm in}}^2\Omega_{\rm in} - \big( {R_{\rm in}}^2 + \frac{\eta_1-\eta_0}{\eta_1+\eta_0}{R_{\rm m}}^2 \big)\Omega_{\rm out} } {1-\frac{{R_{\rm in}}^2}{{R_{\rm out}}^2} +  \frac{\eta_1-\eta_0}{\eta_1+\eta_0}\big(\frac{{R_{\rm in}}^2}{{R_{\rm m}}^2}-\frac{{R_{\rm m}}^2}{{R_{\rm out}}^2}\big)  }, \\ \nonumber
B_1 &=& \frac{  \big( {R_{\rm out}}^2 - \frac{\eta_1-\eta_0}{\eta_1+\eta_0}{R_{\rm m}}^2 \big)\Omega_{\rm in} -\frac{2\eta_0}{\eta_1+\eta_0} {R_{\rm out}}^2\Omega_{\rm out}} {\frac{{R_{\rm out}}^2}{{R_{\rm in}}^2}-1 +  \frac{\eta_1-\eta_0}{\eta_1+\eta_0}\big(\frac{{R_{\rm out}}^2}{{R_{\rm m}}^2}-\frac{{R_{\rm m}}^2}{{R_{\rm in}}^2}\big)  }.
\end{eqnarray}
The coefficients $A_0$ and $A_1$ are obtained by substituting Eq.~(\ref{eq:dw-a}) into Eq.~(\ref{eq:dw_cfa}).

\begin{figure}
\includegraphics{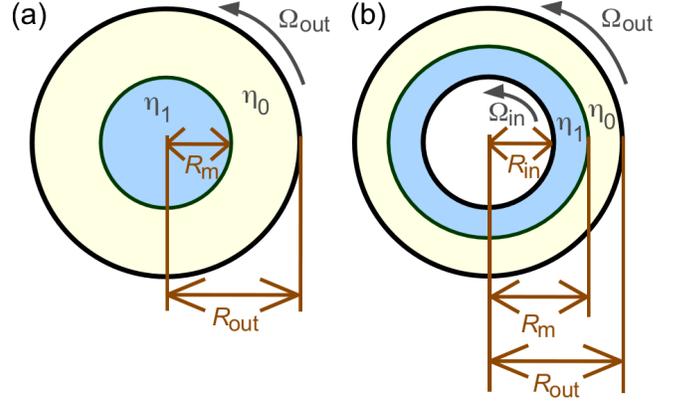}
\caption{
Geometry of (a) a single cylinder and (b) double cylinders for circular laminar flow.
Two fluids with viscosities $\eta_0$ and $\eta_1$ respectively fill
 yellow (white) and blue (gray) regions of the cylinders.
The inner and outer cylinders are rotated with angular velocities $\Omega_{\rm in}$ and $\Omega_{\rm out}$, respectively.
}
\label{fig:cyl}
\end{figure}

\section{Lagrangian finite volume method}\label{sec:LFVM}

\subsection{Angular-Momentum Conservation}\label{sec:AMC}

The finite volume method is widely used to simulate fluids~\cite{ferz02,wend09}.
A fluid is divided into small cells, and the time evolutions of quantities averaged in each cell are considered.
The average gradient is calculated using the surface integral according to the divergence theorem:
\begin{equation}\label{eq:div}
\int_{V_i} \nabla f({\bf r})\ dV = \int_{S_i} f({\bf r})\hat{\bf n}\ dS,
\end{equation}
for a scalar function $f({\bf r})$, where $\hat{\bf n}$ is the unit normal vector of the surface $S_i$ of the $i$-th cell
and points outwards.
A Lagrangian version (LFVM) has been developed for both inviscid fluids~\cite{spri10,duff11} and viscous fluids~\cite{serr01,serr06}.
A Voronoi tessellation is employed in LFVM [see Fig.~{\ref{fig:vor}}(a)].
The mass of the $i$-th cell, $M_i$, is given by
$M_i= \int_{V_i} \rho({\bf r}) dV $.
Since the method is Lagrangian, $M_i$ is constant during the time evolution: $DM_i/Dt=0$.
The mesh-generating point of the $i$-th cell, {\it i.e.}, the vertex of the Delaunay triangulation,
moves with the velocity ${\bf v}_i$.
The velocity evolution is given by integrating Eq.~(\ref{eq:nse0}) as
\begin{equation}\label{eq:FVM0}
M_i \frac{D{\bf v}_i}{Dt} =  \int_{V_i} \nabla {\boldsymbol \sigma}\ dV 
= \int_{S_i} {\boldsymbol \sigma}\hat{\bf n}\ dS.
\end{equation}
The momenta of the neighboring Voronoi cells are transported via the stress on the interface surface between the cells.
The right-hand side gives the forces
to the $i$-th Voronoi cell, such that
Eq.~(\ref{eq:FVM0}) is interpreted as Newton's equation of motion for the center of mass of the fluid in the $i$-th cell.

In substituting the symmetric stress of Eq.~(\ref{eq:strs+a}) into  Eq.~(\ref{eq:FVM0}),
the velocity evolution conserves the angular momentum ({\it i.e.}, $+a$ fluids):
\begin{equation}\label{eq:FVM+a}
M_i \frac{D{\bf v}_i}{Dt} = \int_{S_i}  -P\hat{\bf n} + \eta (\hat{\bf n}\cdot \nabla){\bf v}_i + \eta \nabla v_{\rm n} + \lambda(\nabla \cdot {\bf v})\hat{\bf n} \ dS,
\end{equation}
where  $v_{\rm n}= {\bf v}\cdot \hat{\bf n}$  is the velocity normal to the interface.
In contrast,
the integration of NSE (Eq.~(\ref{eq:nse1})) using  Eq.~(\ref{eq:div})
gives a velocity evolution that DOES NOT conserve the angular momentum:
\begin{equation}\label{eq:FVM-a}
M_i \frac{D{\bf v}_i}{Dt} = \int_{S_i}  -P\hat{\bf n} + \eta (\hat{\bf n}\cdot \nabla){\bf v}_i + \mu(\nabla \cdot {\bf v})\hat{\bf n}  \ dS.
\end{equation}
Because the third term in Eq.~(\ref{eq:FVM+a}) is missing, the stress is asymmetric.
Even when the NSE of the $+a$ fluids is employed, the resultant equation of motion is for the $-a$ fluids  with $\lambda^{(-)}= \eta+\lambda^{(+)}$.
This is because both of the degenerated stress terms are integrated as the symmetric volume-compression stress, {\it i.e.},
the term originated from the latter half of the shear stress ($\eta\partial v_{\beta}/\partial x_{\alpha}$) 
is also treated as the volume-compression stress ($\lambda(\nabla \cdot {\bf v})\delta_{\alpha\beta}$).
Since the former half of the shear stress is only left, the stress becomes asymmetric.
Thus, the loss of AMC is caused by the treatment of the degenerated stress terms.

This mechanism of AMC loss is commonly utilized in computational fluid dynamics (CFD).
The last viscous term, $\mu \nabla(\nabla\cdot{\bf v})$, of NSE is typically implemented as the gradient of the volume stress.
From the viewpoint of the numerical implementation, it is reasonable
to employ  the viscous term that can be discretized more accurately than the other.
Since it is more accurate than separately discretizing both terms,
numerical accuracy is improved and numerical costs are reduced.
In square meshes, the calculation 
of the derivatives along the tangential direction of the cell surface
is as easy as along the normal direction.
However, in non-regular meshes, those along the tangential direction are less accurate.
The above argument for LFVM is straightforwardly applied to FDM, FEM, and SPH.
Effective volumes can be considered for quantities at vertices in FDM,
local variables in shape functions in FEM, quantities for SPH particles.
For SPH, the average in the Voronoi cell is replaced by
the weighting average as $f_i =\sum_j f(r_{ij}) w(r_{ij})$,
where  $r_{ij}$ is the distance between the $i$-th and $j$-th particles and $w(r_{ij})$ is a bell-shaped weight function.
The SPH forces between neighboring particles $f_{ij}w'(r_{ij})$ are interpreted 
as the integral of an immersed surface expressed by $w'(r_{ij})$. 
A similar surface expression can be found in phase field models~\cite{bibe05}.
Since the equation of motion of the SPH particles is derived from the above discretization of Eq.~(\ref{eq:nse1}),
the resultant equation follows Eq.~(\ref{eq:FVM-a}). 
This is the reason for the lack of AMC in conventional SPH,
and AMC is not recovered even if higher-resolution versions are developed in this scheme.

As described in Sec.~\ref{sec:theory},
the equation of motion, Eq.~(\ref{eq:FVM-a}), reproduces the same behavior as Eq.~(\ref{eq:FVM+a}) in bulk fluids;
Artificial flows are generated depending on the boundary conditions on rigid or deformable objects including contact with other fluids.
When the boundary conditions are imposed by the stress balance in Eq.~(\ref{eq:FVM-a}),
the artificial torque on the boundary induces artificial flows as demonstrated for the  circular laminar flows in Sec.~\ref{sec:circular}.
However, in  deterministic simulations without thermal fluctuations,
the boundary condition is more frequently imposed as a separated condition from the bulk velocity evolution
and the symmetric stress of Eq.~(\ref{eq:strs+a}) is calculated from the flow field.
Thus, the boundary is  implemented in accordance with AMC.
In this case, flows of the $+a$ fluids are obtained even when a simulation method lacks AMC.
Thus, the implementation of the boundaries is a significant step in avoiding the artifacts induced by the lack of AMC.
On the other hand, the angular-momentum-conserving ($+a$) methods have an advantage in that
the bulk and boundary fluids are  consistently implemented, such that the boundary conditions are simplified.

In contrast, in particle-based simulation methods involving thermal fluctuations (DPD, MPC, DSMC, and SDPD),
it is difficult to separately impose the stress balance on the boundaries.
The boundary stress can disturb the thermodynamic balance. 
Since the instantaneous flow field is noisy,
a  time average is needed in order to obtain a sufficiently accurate flow field for the stress estimation.
Therefore, the instantaneous stress must be directly calculated from the equation of motion
in order to impose the stress balance on the boundary.
Thus, for these methods, the $+a$ methods are required to simulate the flows with deformable boundaries.

When the viscosity is varied in space and the gradient $\nabla\eta({\bf r})$ or  $\nabla\lambda({\bf r})$ is not negligibly small,
the viscous terms are not  degenerated; thus, 
Eqs.~(\ref{eq:nse0}) and (\ref{eq:FVM+a}) are chosen for NSE and LFVM, respectively.
In this case,  the $+a$ methods are required for both deterministic and stochastic simulations.

\begin{figure}
\includegraphics[width=8.5cm]{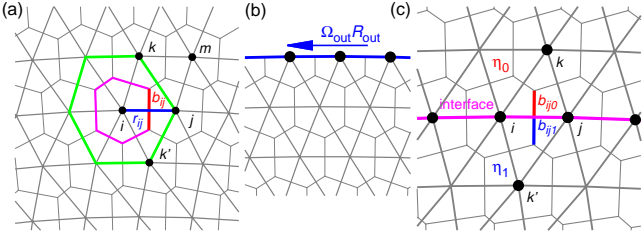}
\caption{
Voronoi tessellation of fluids between two concentric cylinders.
(a) The thick gray (magenta and green) lines of the inner and outer polygons represent the Voronoi cell of the vertex $i$ and 
the nearest neighbor vertices, respectively.
(b) The boundary of the outer cylinder. The closed circles ($\bullet$) represent the boundary vertices.
(c) The interface between inner and outer fluids.
The thick gray (magenta) lines represent the interface.
}
\label{fig:vor}
\end{figure}

\subsection{Discretization of Two-Dimensional LFVM}\label{sec:LFVM2D}

To see more detailed relations,
we investigated the LFVM algorithm for $+a$ fluids in  2D space.
The first (pressure) term in Eq.~(\ref{eq:FVM+a}) is discretized as
\begin{equation}\label{eq:press}
\int_{S_i}  -P\hat{\bf n}\ dS = \sum_j P_j \frac{\partial A_j}{\partial {\bf r}_{i}},
\end{equation}
where the summation is taken over all Voronoi cells including the $i$-th cell.
The area $A_i$ of the $i$-th Voronoi cell is calculated by
$A_i=\sum_j b_{ij}r_{ij}/4$ with the edge length $b_{ij}= (\cot(\theta_{ikj})+\cot(\theta_{ik'j}) )r_{ij}/2$,
where the summation is taken over the neighbor cells of the $i$-th cell and
the angle $\theta_{ikj}=\arccos(\hat{\bf{r}}_{ik}\cdot \hat{\bf{r}}_{jk} )$
[see Fig.~\ref{fig:vor}(a)].
We use ${\bf r}_{ji}={\bf r}_{j}-{\bf r}_{i}$, $r_{ij}=|{\bf r}_{ij}|$, and $\hat{\bf r}_{ji}= {\bf r}_{ji}/r_{ij}$ 
in this paper.
The equation for the inviscid fluids (no viscous interactions)
exactly conserve the translational and angular momenta;
thus, the inviscid term does not influence the present AMC issue.
The derivation of this term is given in Appendix~\ref{app:gra}.
The pressure $P_i$ is the function of the mass density $\rho_i=M_i/A_i$ of the $i$-th Voronoi cell.
We employ $P_i= (\rho_0c^2/2)[(\rho_i/\rho_0)^2-1]$ following Ref.~\cite{elle07},
where $c$ is the speed of sound and $\rho_0$ is the equilibrium mass density.

The second  term in Eq.~(\ref{eq:FVM+a}) is discretized as described in~\cite{serr06}
\begin{equation}\label{eq:stgreen}
 \int_{S_i} \eta (\hat{\bf n}\cdot \nabla){\bf v}_i\ dS = \eta  \sum_j \frac{b_{ij}{\bf v}_{ji}}{r_{ij}},
\end{equation}
where ${\bf v}_{ji}= {\bf v}_j-{\bf v}_i$
and the summation is taken over the neighbor cells.
The third and fourth terms in Eq.~(\ref{eq:FVM+a}) are divided into the differentials along the interface and along the normal direction, $\hat{\bf n}$, as
\begin{eqnarray}\label{eq:stt0}
\int_{S_i} \eta  \nabla v_{\rm n}\ dS &=& \eta  \int_{S_i} \frac{\partial v_{\rm n}}{\partial x_{\rm t}}\hat{\bf t}+  \frac{\partial v_{\rm n}}{\partial x_{\rm n}}\hat{\bf n}\ dS, \\ \label{eq:stv0}
\int_{S_i}  \lambda(\nabla \cdot {\bf v})\hat{\bf n}\ dS &=&  \lambda \int_{S_i} \frac{\partial v_{\rm t}}{\partial x_{\rm t}}\hat{\bf n}+  \frac{\partial v_{\rm n}}{\partial x_{\rm n}}\hat{\bf n}\ dS,
\end{eqnarray}
where   $\hat{\bf t}={\bf e}_z\times \hat{\bf n}$ is the unit vector along the interface and $v_{\rm t}= {\bf v}\cdot \hat{\bf t}$.
Equation~(\ref{eq:stv0})  conserves translational and angular momenta by itself since it consists of the central forces between neighboring cells.
On the other hand, the first term  of Eq.~(\ref{eq:stt0}) does not conserve the angular momentum by itself,
but it is supposed to be canceled along with the non-conserving part of Eq.~(\ref{eq:stgreen}).

Since the differential along $\hat{\bf t}$ has not been taken into account in the previous studies,
we examined several discretization methods.
Among them, the following two give the best and second-best results;
 we refer to them as type I and type II, respectively.
The first term in Eq.~(\ref{eq:stt0}) is discretized as
\begin{eqnarray}\label{eq:stt1}
\big(\int_{S_i} \frac{\partial v_{\rm n}}{\partial x_{\rm t}}\hat{\bf t}\ dS\big)^{\rm type\ I}  &=& \frac{1}{3} \sum_{\rm dela} ({\bf v}_{jk}\cdot \hat{\bf n}_{jk})\hat{\bf r}_{jk}, \\ \label{eq:stt2}
\big(\int_{S_i} \frac{\partial v_{\rm n}}{\partial x_{\rm t}}\hat{\bf t}\ dS\big)^{\rm type\ II}  &=& \sum_{\rm dela} \frac{b_{jk}{\bf v}_{jk}\cdot \hat{\bf n}_{jk}}{r_{im} \hat{\bf n}_{jk}\cdot\hat{\bf r}_{mi} }\hat{\bf r}_{jk}.
\end{eqnarray}
The first term in Eq.~(\ref{eq:stv0}) is discretized as
\begin{eqnarray}\label{eq:stv1}
\big(\int_{S_i} \frac{\partial v_{\rm t}}{\partial x_{\rm t}}\hat{\bf n}\ dS\big)^{\rm type\ I}  &=& \frac{1}{3} \sum_{\rm dela} ({\bf v}_{jk}\cdot \hat{\bf r}_{jk})\hat{\bf r}_{mi}, \\ \label{eq:stv2}
\big(\int_{S_i} \frac{\partial v_{\rm t}}{\partial x_{\rm t}}\hat{\bf n}\ dS\big)^{\rm type\ II}  &=& \sum_{\rm dela} \frac{b_{jk}{\bf v}_{jk}\cdot \hat{\bf r}_{jk}}{r_{im}}\hat{\bf r}_{mi}. 
\end{eqnarray}
In Eqs.~(\ref{eq:stt1})--(\ref{eq:stv2}), 
 the summation is taken over the edges of the Delaunay triangles [the outer thick lines in Fig.~\ref{fig:vor}(a)],
and $\hat{\bf n}_{jk}$ is the unit normal vector of the edge connecting the vertices $j$ and $k$: $\hat{\bf n}_{jk}\cdot {\bf r}_{jk}=0$; $\hat{\bf n}_{jk}$ points outwards: $\hat{\bf n}_{jk}\cdot {\bf r}_{mi}>0$.
When the divergence theorem is directly applied,
the discretizations are slightly different from the type I forms:
the factor $A_i/A_{\rm dela,i}$ appears instead of  $1/3$ in Eqs.~(\ref{eq:stt1}) and (\ref{eq:stv1}),
and $\hat{\bf n}_{jk}$ appears instead of $\hat{\bf r}_{mi}$ in Eq.~(\ref{eq:stv1}),
where   $A_{\rm dela,i}$ is
 the total area of the Delaunay triangles contacting the vertex $i$ [the area surrounded by the outer thick line in Fig.~\ref{fig:vor}(a)].
These modifications are employed to  maintain the translational-momentum conservation and AMC, respectively. 
However, the angular momentum is not exactly conserved in the type I form for Eqs.~(\ref{eq:stgreen}) and (\ref{eq:stt1}).
The type II form is constructed to recover AMC; the torque between the vertices $j$ and $k$ in Eq.~(\ref{eq:stgreen}) 
is canceled by the torque between the vertices $i$ and $m$  in Eq.~(\ref{eq:stt2}).
In a regular lattice in which all Delaunay triangles are equilateral,
the types I and II discretizations coincide.

The second term in Eqs.~(\ref{eq:stt0}) and (\ref{eq:stv0}) is discretized as
\begin{equation}\label{eq:stcen}
 \int_{S_i}  \frac{\partial v_{\rm n}}{\partial x_{\rm n}}\hat{\bf n}\ dS = 
\sum_j \frac{b_{ij}}{r_{ij}}({\bf v}_{ji}\cdot \hat{\bf r}_{ji}) \hat{\bf r}_{ji},
\end{equation}
where the summation is taken over the neighbor cells as in Eq.~(\ref{eq:stgreen}).
The translational momentum is conserved by itself in all of the above discretization terms in
Eqs.~(\ref{eq:press}), (\ref{eq:stgreen}), and (\ref{eq:stt1})--(\ref{eq:stcen}).
On the other hand, the angular momentum is conserved by itself in Eqs.~(\ref{eq:press}) and (\ref{eq:stv1})--(\ref{eq:stcen}),
while the angular momentum is approximately or exactly conserved by the combination of  Eq.~(\ref{eq:stgreen}) with  Eq.~(\ref{eq:stt1})
or Eq.~(\ref{eq:stt2}), respectively.

\subsection{Rigid-Body Rotation}\label{sec:rbr}

When a cylinder or box is filled with a fluid and is slowly rotated as shown in Fig.~\ref{fig:cyl},
the whole fluid rotates with a constant angular velocity like a rigid body: ${\bf v}_{\rm rot}= {\boldsymbol \Omega}\times {\bf r}$.
The pressure gradient is balanced by centrifugation force,
and the viscous terms vanish in NSE: $\nabla^2 {\bf v}_{\rm rot}=\nabla(\nabla\cdot {\bf v}_{\rm rot})= 0$.
However, these vanishments are not trivial in the discretized equations.

In our LFVM scheme,
the viscous forces of Eqs.~(\ref{eq:stv1})--(\ref{eq:stcen}) vanish in the rigid-body rotation,
since ${\bf v}_{ji}\cdot \hat{\bf r}_{ji}=0$.
The forces of Eqs.~(\ref{eq:stgreen}) and (\ref{eq:stt1}) also vanish, 
owing to the divergence theorem and the fact that the divergence of a constant function is null:
$\sum_j b_{ij} \hat{\bf n}=\sum_j b_{ij} \hat{\bf t}=0$.
However, in the type II scheme, the forces of Eq.~(\ref{eq:stt2}) do not vanish in the rigid-body rotation.
This remaining stress destabilizes the constant rotation and induces an artificial flow 
in which the angular velocity depends on the position in the cylinder as demonstrated in Sec.~\ref{sec:results}.
Thus, the  $+a$ fluids  with the type II discretization do not reproduce the rigid-body rotation,
while the angular momentum is strictly conserved.
In contrast, the $+a$ fluids  with the type I discretization reproduce the rigid-body rotation correctly,
while the angular momentum is not strictly conserved.
Although it is better if one method satisfies both conditions,
 we have not found such a discretization scheme.
However, in the type I method, AMC is sufficiently well maintained to reproduce binary flows
(it will be shown in Sec.~\ref{sec:results}) so that the type I method can be employed in applications.

It is surprising that 
the conditions of AMC in the equation of motion  and the rigid-body rotation are different in CFD, unlike in molecular dynamics (MD) simulations.
MD has no velocity-dependent forces, and the rotational invariance of potentials guarantees AMC and the rigid-body rotation.
Since the viscous terms depend on the velocity in CFD,
 invariance is needed to account for global rotational displacement 
and the global rotation of a constant speed.
Therefore, the viscous terms must be independent of the total velocity 
and total angular velocity of the interacting elements (vertices or particles) in CFD.
The $+a$ versions of DPD and MPC satisfy this condition.
The viscous force in DPD is a function of the relative velocity of two DPD particles along the line connecting them: $({\bf v}_{ij}\cdot \hat{\bf r}_{ij})\hat{\bf r}_{ij}$.
The $+a$ types of the MPC collisions are independent of the total velocity and total angular velocity of the MPC particles in each collision cell.
Thus, the viscous interactions of DPD and MPC are independent of the total angular velocity of the interacting particles.
To see this invariance condition more clearly,
let us consider an artificial pairwise force, ${\bf f}_{{\rm rot},ij}=|{\bf v}_{ij}\times \hat{\bf r}_{ij}|\hat{\bf r}_{ij}$, between the $i$-th and $j$-th particles.
This force conserves the translational and angular momenta, since 
${\bf f}_{{\rm rot},ji}=-{\bf f}_{{\rm rot},ij}$ and 
${\bf r}_{ij}\times {\bf f}_{{\rm rot},ij}=0$. On the other hand, the force amplitude is determined by the angular velocity, $|{\bf v}_{ij}\times \hat{\bf r}_{ij}|$, and is varied by the global rigid-body rotation.
Thus, these two conditions must be distinguished.

In conventional SPH, the viscous force proportional to ${\bf v}_{ji}$ between the neighboring $i$-th and $j$-th particles is employed,
and this force is independent of the positions of the other particles.
Therefore, the SPH viscous force does not vanish in the rigid-body rotation; as a result, the rigid-body rotation cannot be reproduced 
except in cases in which SPH particles are initially arranged in a regular lattice.
In SDPD, thermal fluctuations reduce the anisotropic distribution of the particles,
and the deviation from the rigid-body rotation is likely reduced.

\subsection{Implementation at Boundary Wall and Fluid Interface}\label{sec:bd}

We have  explicitly discretized the boundary on a wall and an interface between two fluids,
and the vertices are placed on them as shown in Fig.~\ref{fig:vor}(b) and (c).
Although a mirror image is widely employed to impose the no-slip boundary condition,
the degeneracy of the Delaunay triangles always occurs at a flat boundary,
since the neighboring vertices and their images form an isosceles trapezoid.
To avoid the reduction in the number of the neighbors due to this degeneracy, 
we employ the boundary vertices instead.
Since we investigate flows with steady interfaces,
the interface vertices are not added or removed in the simulations.
If one considers flows accompanied by large interface deformations,
the rearrangements of the interface vertices are required.

The vertices on the inner and outer boundaries of the cylinders are moved along the boundaries with constant velocities 
$\Omega_{\rm in} R_{\rm in}$ and $\Omega_{\rm out} R_{\rm out}$, respectively.
The area of a Voronoi cell on the boundary is approximately half that of the other cells  [see Fig.~\ref{fig:vor}(b)].

To generate a sharp fluid interface, the viscous forces around the interface cells
are carefully implemented.
The interface Voronoi cell is divided into two regions that each contains one types of fluid,
and the viscous force between the interface vertices depends on the area ratio.
The vertices in the different fluids [{\it e.g.}, the vertices $k$ and $k'$ in Fig.~\ref{fig:vor}(c)] 
do not directly interact.
In our scheme, the boundaries are consistently treated with bulk fluids.
When $\eta_1/\eta_0=1$ is inputted, the equation becomes identical to the bulk equation.
The details of the procedure are described in Appendix~\ref{app:interface}.

\begin{figure}
\includegraphics{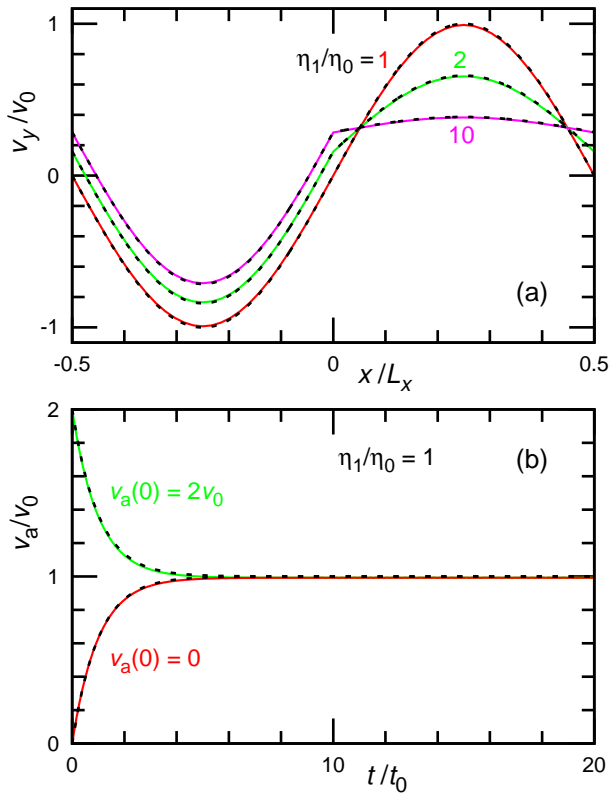}
\caption{
Periodic planar laminar flows.
(a) Velocity profile $v_y(x)$ of the steady states at $\eta_1/\eta_0=1$, $2$, and $10$.
(b) Time evolution of the velocity amplitude $v_{\rm a}$ at $\eta_1/\eta_0=1$.
Two fluids with viscosity $\eta_0$ and $\eta_1$ fill
in the regions $-L_x/2<x\le 0$ and $0<x\le L_x/2$, respectively.
The solid lines are obtained by LFVM for the $+a$ type I method.
The dashed lines represent the analytical solutions of (a) Eq.~(\ref{eq:sin2}) and (b) Eq.~(\ref{eq:sin1}),
and they completely overlap with the solid lines.
The velocity and time are normalized by $v_0=f_0/\eta_0 k^2$
and $t_0=\rho/\eta k^2$, respectively.
}
\label{fig:sin2}
\end{figure}

\subsection{Numerical Details}\label{sec:detail}

In LFVM, it is known that the discretization errors become unacceptably large in disordered meshes~\cite{serr06};
this problem is addressed through rearrangement of the Voronoi vertices~\cite{spri10,duff11}.
Here, we do not employ this rearrangement but
maintain a flow condition in which Voronoi vertices remain in semi-regular arrangements, as shown in Fig.~\ref{fig:vor}.
Since our aim is to clarify the AMC effects, the influence of the  rearrangement is avoided.

Equations (\ref{eq:FVM+a}) and (\ref{eq:FVM-a}) are numerically integrated 
using the fourth-order Runge--Kutta method with $\Delta t= 0.001$.
The parameters $\rho_{\rm int}=\rho_0=1$ and $\eta_0=1$ are used for all flows,
where $\rho_{\rm int}$ is the density in the initial state.
If not specified, $\mu=0$ is used.

For the periodic planar flow,
a simulation box with side lengths of $L_x=52.1$ and $L_y=53.7$ is employed under the periodic boundary condition.
As the initial state, a regular triangular lattice with $A_i=1$ is created, where
the total number of Voronoi cells $N=2800$.
To simulate laminar flows, $f_0=0.01$ and $c=10v_0$ are used, where $v_0=f_0/\eta_0 k^2$.

In the circular laminar flows,
vertices are aligned on concentric circles with mean area $\langle A_i\rangle=1$ as the initial state
and $c=1$ is used.
For the  single cylinder,
$R_{\rm out}=30.1$, $R_{\rm m}=15.1$, and $\Omega_{\rm out}=0.0001$ are used.
For the double cylinders,
$R_{\rm out}=60.1$, $R_{\rm m}=40.1$, $R_{\rm in}=20.1$, and $\Omega_{\rm out}-\Omega_{\rm in}=0.0001$
are used. 
We conducted simulations with other parameter values
to confirm that the results are not sensitive to the above-mentioned parameter choices.

\section{Simulation results}\label{sec:results}

\subsection{Planar Laminar Flows}\label{sec:replanar}

Before showing the AMC effects in circular flows,
we describe the results of the planar laminar flows induced by the periodic force,
and we discuss the accuracy of our simulation methods.

In the single-component fluids,
the steady flow profile and the velocity relaxation from ${\bf v}(0)=0$ or ${\bf v}(0)=2v_0\sin(kx){\bf e}_y$ agree 
well with the analytical predictions 
as shown in Fig.~\ref{fig:sin2}(a) and (b), respectively.
The simulation data overlap with the analytical curves.
The values of the velocity amplitude $v_a$ are calculated by the Fourier transformation of the velocity $v_y$.
Although Fig.~\ref{fig:sin2} shows only the results of the $+a$ fluids with type I discretization at $\mu=0$,
the other combinations of the methods give very similar results
and the differences  only appear in the amplitudes $v_a$.
The viscosity of the simulation is estimated from this amplitude as $\eta_{\rm sin}=f_0/v_ak^2$ 
as shown in Fig.~\ref{fig:vis}.
At $\mu=0$, all methods generate a viscosity close to the input viscosity $\eta_0$.
As $\mu$ increases, the viscosity linearly decreases; moreover, the type II discretization
gives a lower slope than that of type I.
From this aspect, the type II discretization is better.
Compared to it, the effects of AMC ($+a$ or $-a$ fluids) are marginal.
To estimate the numerical errors, we also calculate the viscosity at $f_0=0.005$, where the other parameters are same; The difference of $\eta_{\rm sin}$ is negligibly small: $\Delta \eta_{\rm sin}/\eta_0 <0.002$ for $+a$ type II and  $\Delta \eta_{\rm sin}/\eta_0 <0.0002$ for the other three methods.

Hereafter, we only show the results for $\mu=0$.
The simulation results at $\mu>0$ are very similar to those at $\mu=0$,
and the differences can be understood by rescaling the velocity using $\eta_0/\eta_{\rm sin}$.
For the $-a$ fluids, 
 the choice between the type I or II viscous term does not need to be considered at $\mu=0$,
since they vanish at $\mu=0$.

For two fluids with different viscosities,
the velocity profile has a kink at the fluid interface as shown in Fig.~\ref{fig:sin2}(a).
Again, all of the methods ($+a$ or $-a$ fluids with type I or type II)
show excellent agreement with the analytical solutions.
Thus,  the planar laminar flows are not influenced by the lack of AMC.

\begin{figure}
\includegraphics{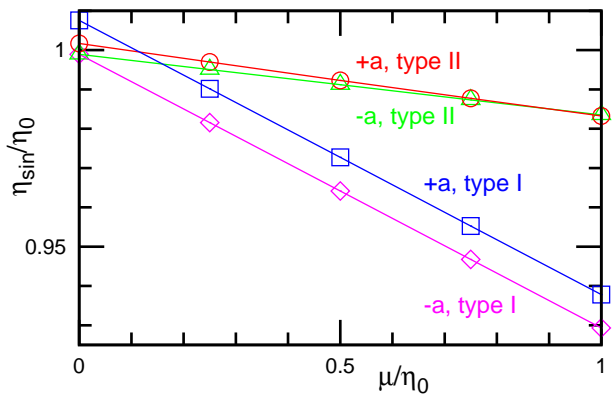}
\caption{
Viscosity $\eta_{\rm sin}$ calculated from the planar laminar flow
for LFVM with 
symmetric ($+a$) and asymmetric ($-a$) viscous terms, using two types of 
discretization methods [type I Eqs.~(\ref{eq:stt1}) and (\ref{eq:stv1}) and type II Eqs.~(\ref{eq:stt2}) and (\ref{eq:stv2})].
The dependence on the coefficient, $\mu$, of the last term in Eq.~(\ref{eq:nse1}) is shown.
}
\label{fig:vis}
\end{figure}

\begin{figure}
\includegraphics{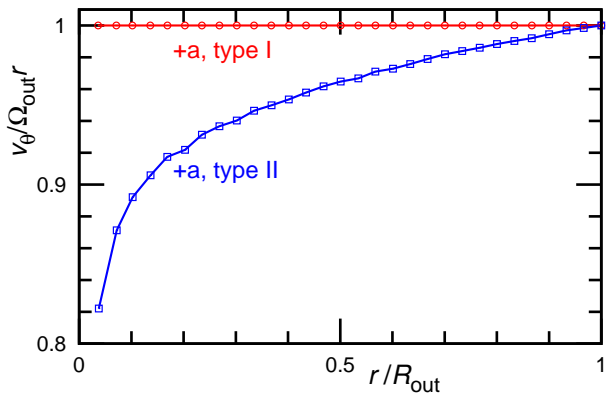}
\caption{
Angular velocity $v_{\theta}/r$ of the single-component $+a$ fluids in the single cylinder.
The fluid using the type I viscous term exhibits the rigid-body rotation,
while the fluid using  the type II term exhibits a non-uniform rotation.
}
\label{fig:sng_dcos}
\end{figure}

\begin{figure}
\includegraphics{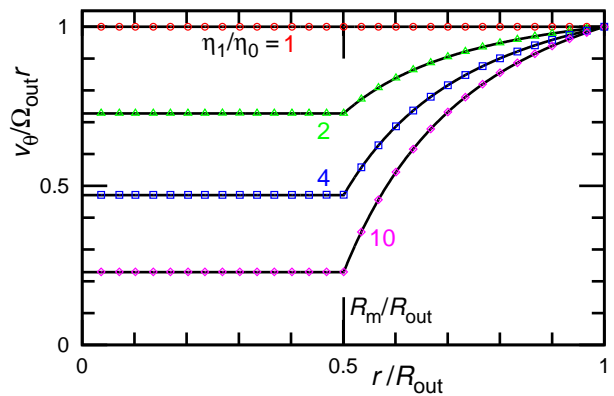}
\caption{
Angular velocity $v_{\theta}/r$ of the binary $-a$ fluids in the single cylinder.
The fluids do not exhibit the rigid-body rotation at $\eta_1/\eta_0 \ne 1$.
The symbols $\circ$, $\triangle$, $\Box$, and $\diamond$ represent
the simulation data for $\eta_1/\eta_0 = 1$, $2$, $4$, and $10$, respectively.
The solid lines represent the analytical solution of Eq.~(\ref{eq:v_cf}) with Eq.~(\ref{eq:sng-a}).
}
\label{fig:sng-a}
\end{figure}

\begin{figure}
\includegraphics{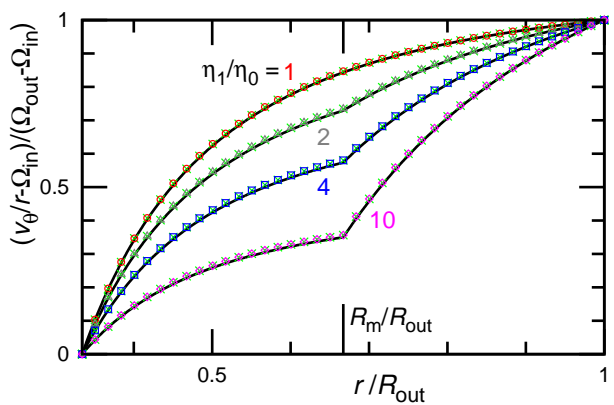}
\caption{
Angular velocity $v_{\theta}/r$ of the binary $+a$ fluids in the double cylinders.
The solid lines represent the analytical solution of Eq.~(\ref{eq:v_cf}) with Eqs.~(\ref{eq:dw+a}) and (\ref{eq:dw_cfa}).
The data ($\times$) for $\Omega_{\rm out}=0$ and $\Omega_{\rm in}=-0.0001$ overlap 
with those for $\Omega_{\rm out}=0.0001$ and $\Omega_{\rm in}=0$ 
($\circ, \triangle, \Box$, and $\diamond$ for $\eta_1/\eta_0 =1, 2, 4$, and $10$, respectively).
}
\label{fig:dw+a}
\end{figure}

\begin{figure}
\includegraphics{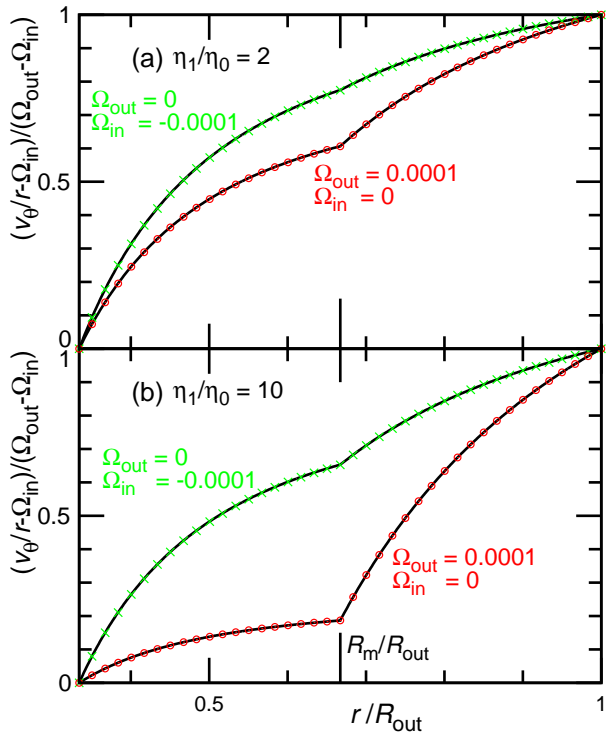}
\caption{
Angular velocity $v_{\theta}/r$ of the binary $-a$ fluids in the double cylinders 
for (a) $\eta_1/\eta_0 = 2$ and (b) $\eta_1/\eta_0 = 10$.
The symbol $\circ$ ($\times$) represents
the simulation data at $\Omega_{\rm out}=0.0001$ and $\Omega_{\rm in}=0$ ($\Omega_{\rm out}=0$ and $\Omega_{\rm in}=-0.0001$).
The solid lines represent the analytical solution of Eq.~(\ref{eq:v_cf}) combined with Eqs.~(\ref{eq:dw_cfa}) and (\ref{eq:dw-a}).
}
\label{fig:dw-a}
\end{figure}

\subsection{Circular Laminar Flows}\label{sec:recircular}

First, a single-component fluid filling a single cylinder is considered.
The type I $+a$ fluid and the $-a$ fluid exhibit a rigid-body rotation;
the angular velocities are constant (see the upper lines in Figs.~\ref{fig:sng_dcos} and \ref{fig:sng-a}).
However, the type II $+a$ fluid exhibits an artificial flow as described in Sec.~\ref{sec:rbr}
(see the lower line in Fig.~\ref{fig:sng_dcos}).
Therefore, the invariance of the viscous forces under the global rotation
is significant in the simulation of rotational flows.
For this reason, the type I method is better than the type II method,
although the type II method provides better viscosity estimation for $\mu>0$.
The invariance under the global rotation is more important than
the exact AMC in the equation of motion.
The type I method maintains AMC not exactly but sufficiently well
to reproduce binary flows as shown below.
Hereafter, we use only the type I method for the $+a$ fluid.

The type I $+a$ binary fluids maintain constant angular velocity
 in a single cylinder (see Appendix~\ref{app:interface}).
In contrast, in the $-a$ binary flow, the outer fluid exhibits
the angular-velocity gradients that are well predicted by the analytical solutions in Sec.~\ref{sec:circular},
as shown in Fig.~\ref{fig:sng-a}.

The results from the simulations with double cylinders
also have excellent agreement with the analytical solutions for both $+a$ and $-a$ fluids, as shown in Figs.~\ref{fig:dw+a} and \ref{fig:dw-a}.
In the $+a$ fluids, the velocity gradient is independent of the absolute angular velocity;
the simulation results of the outer-wall rotation ($\Omega_{\rm out}=0.0001$ and $\Omega_{\rm in}=0$) 
 completely overlap with those of the inner-wall rotation ($\Omega_{\rm out}=0$ and $\Omega_{\rm in}=-0.0001$).
On the other hand, the velocity gradient of the $-a$ fluids depends on the absolute angular velocity.
Thus, the AMC of the viscous interactions is significant in simulating multi-component fluids with different viscosities.

\section{Summary and discussion}\label{sec:sum}

We have clarified the conditions of angular-momentum conservation and the invariance
under global rotation in fluid dynamics simulations. In NSE, the viscous terms for volume
change and half of the shear deformation are degenerated for a fluid with a constant
viscosity, although they are the gradients of symmetric and asymmetric stresses,
respectively. To maintain the AMC of the discretized equation of motion, these degenerated
terms must be separately implemented in accordance with their stress origins. If the
degenerated terms are implemented together as the gradient of the symmetric stress, the
resultant forces for the shear deformation lack AMC. On the other hand, the invariance
of the viscous forces in a global rotation with a constant speed is a necessary condition for
reproducing the rigid-body rotation of a fluid in a slowly rotating container.

When applying CFD, the torque balance on wall boundaries and fluid interfaces
is essential in reproducing flows involving a rotation. When the boundary implementation
lacks AMC, multi-component fluids with different viscosities generate
artificial flows. To avoid them, 
one has the following two choices.
(i) A $+a$ CFD method is employed for all of the fluids.
(ii) The bulk fluids are solved by a $-a$ method but 
the boundary fluids are solved by the stress balance where the symmetric viscous stress
is calculated from the velocity field.

To calculate low Mach number flows of liquids, the incompressible assumption is widely
employed. In incompressible fluids, the last term of NSE is removed ($\mu=0$) and the
incompressible constraint is imposed. Thus, the stress becomes asymmetric. However, the
incompressible constraint implicitly transports momentum
 and a fluid element can interact with itself via the periodic boundary.
Although it originates from a symmetric stress (pressure),
 it is not completely obvious whether it loses AMC. This is an important
issue that should receive further study.

We developed LFVM to demonstrate the effects of AMC in binary fluids. The
discretizations of the viscous stresses and the fluid interfaces were newly implemented. The
simulation results show excellent agreement with the analytical solutions. In developing
other fluid dynamics methods, the steady rotational flows presented in this study are
suitable for checking whether a method reproduces AMC flows.

The LFVM implementation in this study is difficult to directly apply to SPH,
since the SPH interaction is constructed in a  pairwise manner.
In the $+a$ version of SDPD~\cite{mull15}, the spin variable is introduced.
Instead of imposing the asymmetric stress in the equation of motion,
the asymmetric viscous force is stored in the spin velocity
and released with a time delay.
It can be interpreted that the lacked asymmetric stress is implicitly
imposed via the spin rotation.

\begin{acknowledgments}
This work was supported by JSPS KAKENHI Grant Number JP17K05607 
and MEXT as ``Exploratory Challenge on Post-K computer'' (Frontiers of Basic Science: Challenging the Limits).
\end{acknowledgments}

\begin{appendix}

\section{Gradient Discretization}\label{app:gra}

The volume integral of the gradient in a Voronoi cell 
is expressed by the surface integral of the cell, as described in Eq.~(\ref{eq:div}).
When the surface value is approximated as the mean value of two contacting cells, $f_s= (f_i+f_j)/2$, 
the gradient is given by
\begin{equation}\label{eq:gra1}
(\nabla f)_i =  \frac{1}{V_{i}}\int_{V_i} \nabla f({\bf r})\ dV = \frac{1}{V_{i}}\sum_j \frac{S_{ij}}{2}(f_i+f_j) \hat{\bf r}_{ji},
\end{equation}
where $V_i$ is the volume (area in 2D) of the $i$-th Voronoi cell
and $S_{ij}$ is the area (length in 2D) of the interface between the $i$-th and $j$-th Voronoi cells.
The summation is taken over the cells neighboring the $i$-th cell.
However, this discretization is inaccurate.
To improve the accuracy,
Serrano {\it et al.}~\cite{serr05} and Springel~\cite{spri10} independently derived
the following form:
\begin{equation} \label{eq:gra2}
(\nabla f)_i = \frac{1}{V_{i}} \sum_j S_{ij}\Big[(f_i+f_j) \frac{\hat{\bf r}_{ji}}{2} + (f_j-f_i)\frac{{\bf c}_{ij}}{r_{ij}}\Big].
\end{equation}
The vector ${\bf c}_{ij}$ of the center of the interface
from the middle position $({\bf r}_i+{\bf r}_j)/2$ between the $i$-th and $j$-th Voronoi cells is given by
\begin{equation}\label{eq:cij}
{\bf c}_{ij} = \frac{1}{S_{ij}}\int_{S_{ij}} \Big( {\bf r} - \frac{{\bf r}_i+{\bf r}_j}{2} \Big)\ dS.
\end{equation}
Although Eq.~(\ref{eq:gra2}) works quite well,
the second term does not conserve the angular momentum,
whereas it conserves the translational momentum.

\begin{figure}
\includegraphics{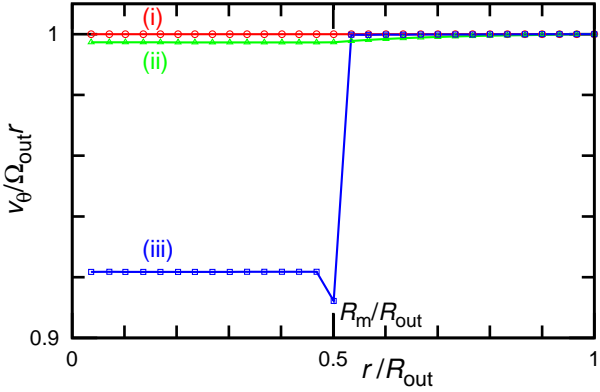}
\caption{
Angular velocity $v_{\theta}/r$ of the binary $+a$ fluids in the single cylinder at $\eta_1/\eta_0 =10$.
Three implementations representing the viscosity stress on the fluid interface are compared for the type I discretization.
}
\label{fig:sng_trav}
\end{figure}

We introduce an AMC discretization form:
\begin{equation} \label{eq:gra3}
(\nabla f)_i = - \frac{1}{V_{i}} \sum_j  f_j\frac{\partial V_j}{\partial {\bf r}_i},
\end{equation}
where the summation is taken over all cells.
Equation~(\ref{eq:gra3}) is derived from the Lagrangian equation for the pressure in Ref.~\cite{serr05} and from 
$\int \partial f(r)/\partial {\bf r}_i dV=0$ and $\int f(r) dV= \sum_i f_{i}V_i$ in Ref.~\cite{spri10}.
Nevertheless,  Eq.~(\ref{eq:gra2}) was employed instead, since
Eq.~(\ref{eq:gra3}) was considered to be identical to Eq.~(\ref{eq:gra2}).
However, they are not exactly equal.
Equation~(\ref{eq:gra3}) conserves the translational and  angular momenta, 
since $V_i$ is invariant for the translation and rotation of the coordinates, unlike  Eq.~(\ref{eq:gra2}).
The first and second terms in Eq.~(\ref{eq:gra2}) represent the translation and rotation of the interface $S_{ij}$, respectively;
thus, the change in the interface area (length in 2D) is not taken into account.

\section{Viscous Interactions on Fluid Interface}\label{app:interface}

The type I viscous forces are modified on the fluid interface as follows.
The viscous forces along the interface [between the vertices $i$ and $j$ in Fig.~\ref{fig:vor}(c)] are calculated 
using Eqs.~(\ref{eq:stgreen}) and (\ref{eq:stcen}) with
\begin{equation} \label{eq:etam}
\eta= \frac{b_{ij0}\eta_0+b_{ij1}\eta_1}{b_{ij}},
\end{equation}
and $\lambda$ is similarly averaged, where the Voronoi edge is divided into two regions, each with one type of fluid: 
$b_{ij0}= \cot(\theta_{ikj})r_{ij}/2$ and $b_{ij1}= \cot(\theta_{ik'j})r_{ij}/2$ [see Fig.~\ref{fig:vor}(c)].
For the viscous forces of Eqs.~(\ref{eq:stt1}) and (\ref{eq:stv1}),
the viscosity of the Delaunay triangle $ijk$ is employed.
Since the stresses $\sigma_{kij}$ and $\sigma_{k'ij}$ on the vertices $k$ and $k'$ for the triangles $ijk$ and $ijk'$  in Fig.~\ref{fig:vor}(c) 
are not balanced at $\eta_1\ne \eta_0$, 
the remaining stresses are imposed on the interface vertices $i$ and $j$: $ \sigma_{i}^{\rm fi}=\sigma_{j}^{\rm fi}= -(\sigma_{kij}+\sigma_{k'ij})/2$.
This means that the edge $ij$ exerts stress on the vertex $k$ and receives the opposite stress by Newton's third law.
Using these implementations, 
the velocity fields around the interface are properly reproduced.
The binary fluids with $\eta_1/\eta_0=10$ exhibit constant-speed rotation,
as shown in line (i) in Fig.~\ref{fig:sng_trav}.

To clarify the improvements provided by the above implementation, we examined the following two modifications [(ii) and (iii)].
(ii) $\eta= (\eta_0+\eta_1)/2$ is employed instead of Eq.~(\ref{eq:etam}):
The velocity of the inner fluid slightly deviates as shown in line (ii) in Fig.~\ref{fig:sng_trav},
since the area of the outer fluid is larger than $1/2$ on convexly curved interfaces.
(iii) The average viscosity of Eq.~(\ref{eq:etam}) is employed for the stresses on the vertices $k$ and $k'$ 
to balance $\sigma_{kij}$ and $\sigma_{k'ij}$:
A gap in the angular velocity appears at the interface as shown in line (iii) in Fig.~\ref{fig:sng_trav}.
Since the type II stress term involves the interaction between $k$ and $k'$,
the type II fluids exhibit a similar gap at the interface.

Thus, the correct implementation of the interface is important in reproducing the flow behavior 
involving a zero or negligibly small angular-velocity gradient. 
For flows involving the angular-velocity gradient such as the  circular Couette flows shown in Fig.~\ref{fig:dw+a},
this gap is significantly reduced.

\end{appendix}

\end{document}